\documentclass[reprint,onecolumn,amsmath,amssymb,jap,aip]{revtex4-1}

\usepackage{graphicx}
\usepackage{dcolumn}
\usepackage{bm}
\usepackage{epsfig}
\usepackage{mathptmx}
\usepackage{times}
\usepackage{amsmath}
\usepackage{amssymb}
\usepackage{dcolumn}
\usepackage{units}
\usepackage{upgreek}
\usepackage{tabularx}
\usepackage{todonotes}
\usepackage{threeparttable}
\usepackage[space]{grffile}
\usepackage[normalem]{ulem}

\usepackage{subfig}
\usepackage{multirow}
\usepackage{array}
\usepackage{xcolor}
\usepackage{soul}
\usepackage[]{caption}
\usepackage{physics}
\usepackage[T1]{fontenc}
\usepackage{units}
\usepackage{tgschola} %utopia

\begin{document}

\title{A device-level compact model for mushroom-type phase change memory}
\author{Stephan Menzel}\affiliation{Peter Gruenberg Institut (PGI-7), Forschungszentrum J\"{u}lich, Germany}\affiliation{IBM Research -- Europe, S\"{a}umerstrasse 4, 8803 R\"{u}schlikon, Switzerland} %\email{st.menzel@fz-juelich.de,ghs@zurich.ibm.com}
\author{Benedikt Kersting}\affiliation{IBM Research -- Europe, S\"{a}umerstrasse 4, 8803 R\"{u}schlikon, Switzerland}
\author{Rana Walied Ahmad}\affiliation{Peter Gruenberg Institut (PGI-7), Forschungszentrum Juelich, Germany}
\author{Abu Sebastian}\affiliation{IBM Research -- Europe, S\"{a}umerstrasse 4, 8803 R\"{u}schlikon, Switzerland}
\author{Ghazi Sarwat Syed}\affiliation{IBM Research -- Europe, S\"{a}umerstrasse 4, 8803 R\"{u}schlikon, Switzerland}

\maketitle

% Abstract 
\noindent{\textbf{In this work we introduce a compact model for mushroom-type phase-change memory devices that incorporates the shape and size of the amorphous mark under different programming conditions, and is applicable to both projecting and non-projecting devices. The model includes analytical equations for the amorphous and crystalline regions and uniquely features a current leakage path that injects current at the outer edge of the electrodes. The results demonstrate that accurately modeling the size and shape of the phase configurations is crucial for predicting the full-span of the RESET and SET programming, including the characteristics of threshold switching. Additionally, the model effectively captures read-out behaviors, including the dependence of resistance drift and bipolar current asymmetry behaviours on the phase configurations.  The compact model is also provided in Verilog-A format, so it can be easily used in standard circuit-level simulation tools.}}

\begin{flushleft}
 Keywords: Phase Change Memory, Threshold Switching, Compact Modeling
\end{flushleft}

\section*{Introduction}

\noindent The ability to capture the complex and dynamic behavior of functional nanoscale devices through precise mathematical models facilitates enhanced and fast-paced device optimization and engineering. Such models become even more relevant for analyzing highly non-linear memristive systems, such as phase-change memory (PCM), in which both static and dynamically changing state variables determine the device behavior. These devices are candidates for data storage and computing-in-memory applications due to their non-volatility, good retention properties, and compatibility with back-end-of-line CMOS processing \cite{syed2025phase,Burr2016003,Ehrmann2022001,Gallo2020001}. A PCM device, such as the mushroom-type, can be reversibly programmed to a continuum of resistance states in the nanoseconds time scale by exploiting the rich electro-thermal coupling in the carrier transport physics\cite{Nardone2012001,Kaes2015001,Sarwat2022001,Gallo2015001} and crystal growth dynamics \cite{Zhang2019008,Sebastian2014001,Menzel2015001}.  The resistance states are encoded in the phase configuration, viz, the shape and size of the amorphous volume within the crystalline phase change material (PC material). Crucially, the phase configuration is a critical parameter and determines the key device characteristics, such as the threshold voltages \cite{Gallo2016001,Wimmer2014001}, the resistance drift \cite{Kim2011030,Kersting2020001}, and the retention characteristics \cite{Rizzi2014001002,Gallo2020001}. PCM devices have been modeled using a variety of techniques, including physical models, empirical models, and compact models \cite{Tabatabaei2017001,Ciocchini2015001, Woods2017003,Woods2017004,Pigot2018001,Pigot2018002,Ding2022001}. Of particular interest are compact models, which are simplified mathematical representations of the physical behavior of a device. They are relevant since their use can be naturally extended in circuit simulations, where they can be easily integrated into larger systems. \\

Current compact models of PCM devices, however, lack information on phase configurations, using only the fraction of the phase configuration as the state variable, i.e., the fully crystalline (SET) or fully amorphous (RESET) states \cite{Pigot2018001,Pigot2018002,Ding2022001}. Although such models have been shown to capture the essential electrical and thermal properties of the device, they do not take into account the shape of the amorphous dome during programming, which we discuss to be a key parameter. Note that a compact model that considers the shape of the amorphous phase, as observed in continuum simulations, is currently missing. Alternatively, physical continuum simulations based on numerical finite element methods can capture phase configurations, but they are computationally complex and time-consuming, making them unsuitable for circuit simulations \cite{Woods2017003,Woods2017004,Ciocchini2015001}.\\

This study presents a compact model specifically designed for mushroom-type PCM devices, or devices where the amorphous volumes are shaped as domes. The model takes into account the size and shape of the amorphous mark in different programming conditions and is applicable to both unprojected and projected devices. The equations derived in this model account for the shapes of both, the amorphous and crystalline regions. Another important aspect of the model is the inclusion of a phase-configuration-dependent leakage current path that injects current directly at the outer edge of the heater. We find that incorporating these features allows the model to accurately reproduce experimental programming curves and is crucial for correctly predicting threshold voltages, resistance drift and bipolar current asymmetry during read-out. To facilitate easy integration with standard circuit design tools such as SPICE simulators, we also developed a Verilog-A model. Although this work specifically targets mushroom-type devices, the modeling approach presented provides a framework for optimizing and designing PCM devices more broadly.

\section*{An Understanding of Programming Curves}

\noindent A certain amount of programming power is required to amorphize the PC material as the material needs to be heated up above the melting temperature. Multilevel RESET states can be achieved within a small power window, while for higher programming powers the resistance of the full RESET states saturates \cite{Sarwat2022001,Burr2016003}. The programmed higher resistance states show resistance drift to higher resistances due to a logarithmic increase in the resistivity of the amorphous volume over time \cite{Kersting2021001}. In addition to resistance drift behavior, PCM mushroom devices also show an asymmetry in the $I-V$ characteristics with respect to voltage polarity, which has been attributed to the resistance of the interface between the amorphous and crystalline volume, as well as between the amorphous volume and heater \cite{Sarwat2022001}. The amount of asymmetry depends on the programmed resistance state. Essentially, the phase configuration, which represents the shape and size of the amorphous region, is the key property that influences device characteristics and must be accounted for in the simulations. \\

\begin{figure}
  \includegraphics[width=\linewidth]{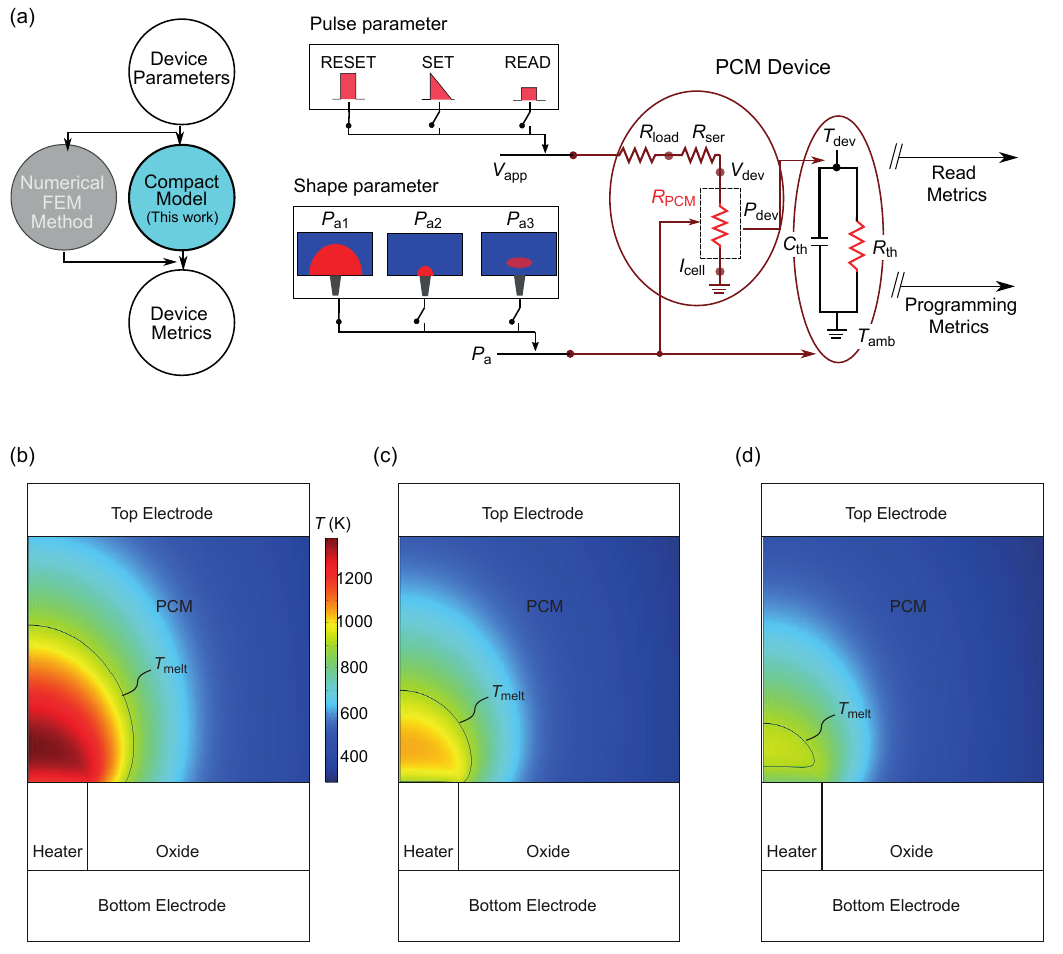}
  \caption{(a) An illustration of the modeling approach and electro-thermal framework of a PCM device, accounting for the phase configurations and measurement-type. (b-d) Simulated temperature distribution during RESET operation under (b) high-power electrical pulse, creating phase configuration Pa1; (c) intermediate pulse power, creating Pa2; and (d) low programming power, creating Pa3. The black solid lines mark the melting temperature of GST phase-change material, $T_\text{m} = 880\,\text{K}$}.
  \label{fig:figure1}
\end{figure}

We address these in our proposed approach (see Figure \ref{fig:figure1}a). Based on continuum simulations, we derive an analytical model. %that can be implemented in SPICE simulators. 
The model can account for various possible phase-configurations: from amorphous marks within the crystalline matrix to a large half-dome shaped amorphous mark covering the heater (bottom electrode). Depending on the programming scheme/programming power, the model can select the most suitable phase configuration and the corresponding equivalent circuit diagram to describe the electrical-thermal characteristics of the device. Briefly, the electrical part of the model describes the current $I_\text{PCM}$ in dependence on the size of the amorphous mark $u_\text{a}$, the temperature $T$, and the electric field $E$ as 

\begin{equation}
    I_\mathrm{PCM} =  f(u_\text{a},T,E).
    \label{equ:Ipcm}
\end{equation}

The electrical model is coupled to a time-dependent thermal model of the device that calculates the device temperature $T_\mathrm{dev}$ as the sum of the ambient temperature $T_\mathrm{amb}$ and the increase in hot spot's transient temperature $T$ due to Joule heating, as described by  

\begin{equation}
   \frac{\mathrm{d}T}{\mathrm{d}t} =  \frac{R_\mathrm{th}I_\mathrm{PCM}V_\mathrm{app}-T}{ R_\mathrm{th}C_\mathrm{th}}
    \label{equ:Tpcm}
\end{equation}

where $R_\mathrm{th}$ is the thermal resistance, and $C_\mathrm{th}$ is the thermal capacitance.\\

As a first step, we calibrate such a model using the temperature distribution (see Figure \ref{fig:figure1}(b)-(d)) in the PCM device during the RESET programming (the details are given in the supplementary information section 1). The sizes of the amorphous marks are related to the isotherm equivalent to the melting temperature $T_\text{melt}$. Based on the amplitude of the RESET pulse power, three general temperature distributions evolve.  For high programming power, the heater is completely covered with the amorphous phase, which shows a dome shape as shown in Figure\,1(b). Later, this configuration is called a homogeneous dome. For one distinct programming power, the heater is exactly covered by a dome-shaped amorphous mark with radius $u_\text{a}=r_\text{heater}=r_\text{be}$ (see Figure\,1(b)). If the heating power is low, the amorphous mark evolves above the heater within the PCM layer, not touching the heater anymore, as shown in Figure \ref{fig:figure1}(c). This configuration will be called a heterogeneous dome.\\

\begin{figure}
  \includegraphics[width=\linewidth]{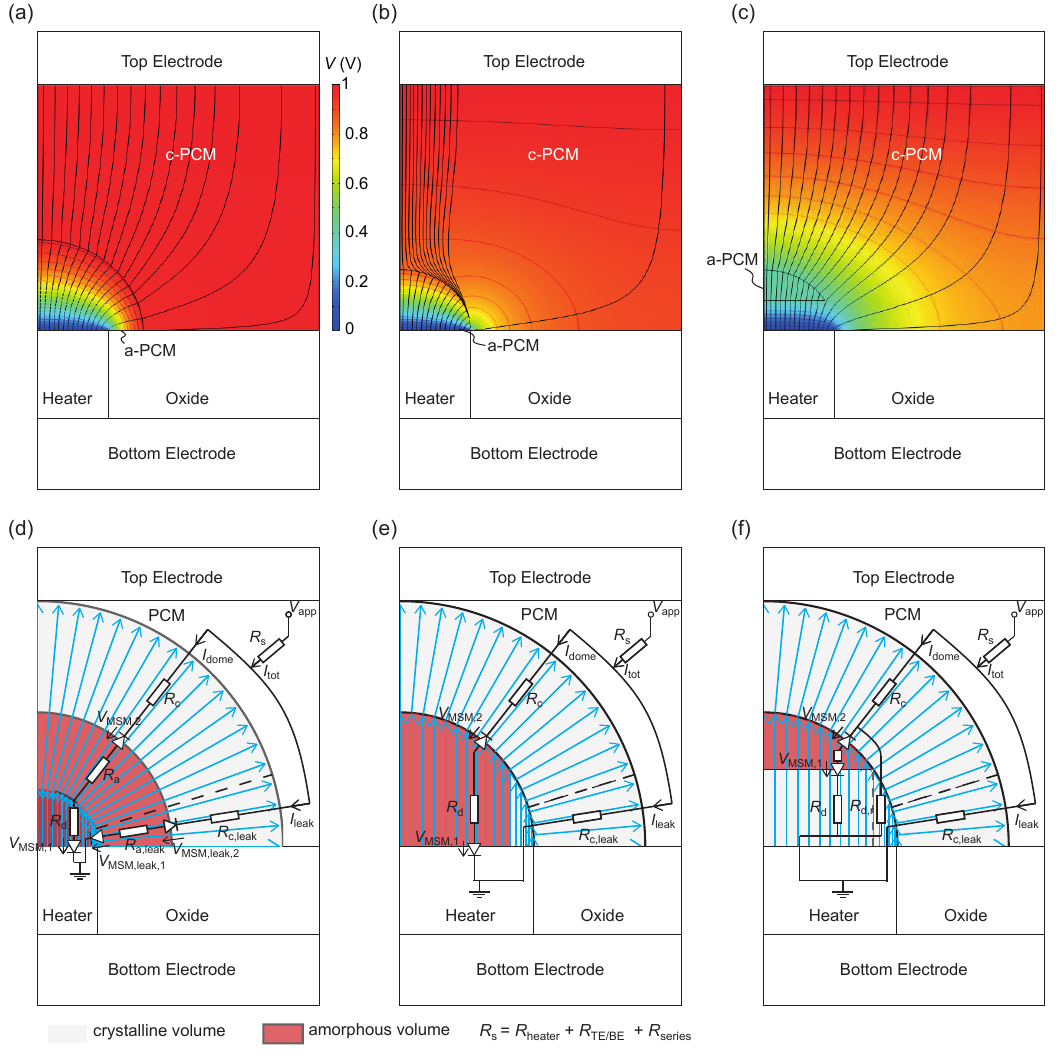}
  \caption{Electric field lines at a voltage of 1~V for (a) a large homogeneous dome  $u_\text{a} > r_\text{be}$, (b) the homogenous amorphous dome with $u_\text{a} = r_\text{be}$, and (c) the heterogeneous dome with $u_\text{a} < r_\text{be}$. The corresponding equivalent circuit diagrams of (d) a larger sized dome $u_\text{a} > r_\text{be}$, (e) the intermediate sized dome with $u_\text{a} = r_\text{be}$, and (f) the  partial RESET state $u_\text{a} < r_\text{be}$.}
  \label{fig:figure2}
\end{figure}

To derive an analytical model for the three different configurations, the electric field lines are analyzed. To this end, the current continuity equation has been solved in the phase change layer for different sizes of the amorphous mark. For simplicity, the amorphous mark is approximated as a hemisphere with radius $u_\text{a}$ as long as $u_\text{a} > r_\text{be}$. If $u_\text{a} <  r_\text{be}$, the amorphous mark does not touch the bottom electrode. In this case, $u_\text{a}$ represents the rightmost point of the amorphous mark, which has the coordinate $(r,z) \left(u_\text{a}, \sqrt{r_\text{be}^2-u_\text{a}^2} \right)$. Thus, the lower boundary of the amorphous mark is $z= \sqrt{r_\text{be}^2-u_\text{a}^2}$. More details of this simulation are given in the supplementary information section 3. The resulting equipotential lines and current streamlines are shown in Figure \ref{fig:figure2}(a)-(c). In addition, the potential distribution is encoded in color. The current streamlines indicate a hypothetical trace of electrons that are injected at the heater and travel to the top electrode. The electric field is directed along these streamlines. The simulation results show three important observations:
(i) The electric field lines in the dome region ($r = r_\text{be}$) point mainly in z-direction. (ii) Outside the dome region ($r > r_\text{be}$), the field lines point mainly in the normal (radial) direction of the dome surface. (iii) There is a significant electric field at ($z=0$) in radial direction for $r_\text{be} < r < {u_\text{a}}$ in Figure\,\ref{fig:figure2}(a). Similar high fields are also observed in this region in Figure\,\ref{fig:figure2}(b) and (c) although the material is crystalline. 
Based on observations (i)-(ii), we approximate the field lines to point in $z$-direction within the dome with ($r=r_\text{be}$), while they point in radial direction outside this region. Observation (iii) led us to introduce a current sneak (leakage) path that describes the current that is injected at the edge of the heater. The resulting equivalent circuit diagrams for the three different configurations are shown in Figures\,\ref{fig:figure2}(d)-(f).\\

For the large dome as shown in Figure\,\ref{fig:figure2}(d), the main current path consists of the resistance $R_\text{d}$ of the amorphous dome, describing the resistance of the hemispherical dome with $r= r_\text{be}$, the resistance $R_\text{a}$ of the amorphous shell, and the resistance $R_\text{c}$ of the crystalline shell. In addition, the heater/amorphous interface and the amorphous/crystalline interface are modeled using diodes with voltage drops $V_\text{MSM,1}$ and $V_\text{MSM,2}$, respectively. The parallel leakage path bypasses $R_\text{d}$ and thus consists only of the diodes with voltage drops $V_\text{MSM,leak,1}$ and $V_\text{MSM,leak,2}$, and the resistances $R_\text{a,leak}$ and $R_\text{c,leak}$ of the amorphous and crystalline shell, respectively. A series resistance $R_\text{s}$, which combines the resistance of the electrodes, the heater, and an external current-limiting resistor, is connected in series to the two parallel current paths. When the amorphous dome shrinks, the volume of the shell resistances decreases until these components vanish when $u_\text{a}$ equals $r_\text{be}$. For this case, the equivalent circuit diagram is shown in Fig.\,\ref{fig:figure2}(e). As the amorphous shell gets vanished, the elements $R_\text{a,leak}$ and the diodes in the leakage path are removed. \\

For the partial RESET state, i.e., the heterogeneous dome, the leakage path consists only of the crystalline shell resistance. As the amorphous region does not touch the heater anymore, the current can bypass the amorphous region, flowing through the crystalline dome region into the heater. Thus, the main current path in the dome region has two parallel branches. The first (right) branch describes the current bypassing the amorphous region and is modeled by resistance $R_\text{d,r}$. The left branch comprises the two diodes modeling the amorphous/crystalline interfaces and the dome resistance $R_\text{d,l}$ that includes the resistance of the amorphous and crystalline region.\\

\begin{figure}
 \includegraphics[width=\linewidth]{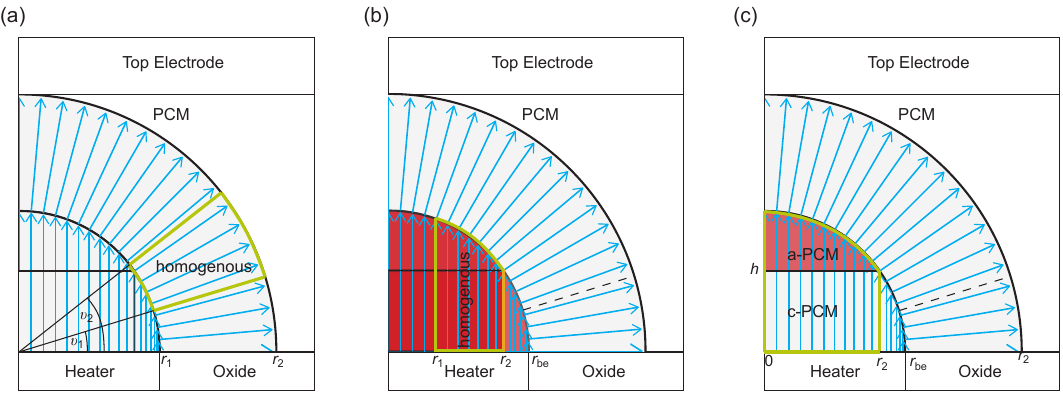}
  \caption{Simplified field distribution and illustration of the resistor elements as marked in green. (a) Shell element bounded by the radii $r_1$ and $r_2$ and the angles $\vartheta_1$ and $\vartheta_2$. (b) Homogenous dome element that is bounded by radii $r_1$ and $r_2$, the sphere $r=r_\mathrm{be}$ and the PC material/heater interface. (c) Heterogenous dome element consisting of a lower crystalline phase change material (c-PCM) and and upper ($z \geq h$ amorphous phase change material part. The element extends laterally from 0 to $r_2$ and is bounded by the the sphere $r=r_\mathrm{be}$ and the PC/heater interface.}
  \label{fig:figure3}
\end{figure}

Based on the simplified field distribution, analytical equations for the resistor elements of the equivalent circuit diagrams can be derived. Based on the shape of the resistors, three different cases can be distinguished as illustrated in Figure\,\ref{fig:figure3}(a)-(c). In the most general case, the shell resistance $R_\text{shell}$ describes the current flow through a shell segment bounded by the angles $\vartheta_1$ and $\vartheta_2$ as well as the radii $r_1$ and $r_2$ (see Figure\,\ref{fig:figure3}(a)). The shape-dependent equation can be derived by integrating along the radial axis as the electric field lines point in this direction (for details, see supplementary information section 4). The integration yields

\begin{equation}
    R_\mathrm{shell} =  \frac{\rho \left( E,T \right)}{2 \pi \left( \cos \left(\vartheta_2 \right) - \cos \left(\vartheta_1 \right) \right)}  \left( \frac{1}{r_2} - \frac{1}{r_1} \right)
    \label{equ:Rshell}
\end{equation}

where $\rho$ is the resistivity.
The second shape configuration is a homogeneous part of the dome that is bounded by the radii $r_1$ and $r_2$, and in $z$-direction by the heater at the bottom and the hemisphere $r=u_\mathrm{a}$ at the top as shown in Fig.\,\ref{fig:figure3}(b). The resistance $R_\mathrm{dome,hom}$ can be derived by considering a parallel configuration of thin cylindrical rings and integrating the conductance of each ring in $r$-direction at $\vartheta = \pi/2$, i.e., along the heater surface. The complete derivation is given in the supplementary information section 4, which results in

\begin{equation}
    R_\mathrm{dome,hom} =  \frac{\rho \left(E,T \right)} {2\pi r_\mathrm{be}} 
    \label{equ:Rdome}
\end{equation}

In the configuration shown in Figure\,\ref{fig:figure3}(c), the material is inhomogeneous consisting of a cylindrical crystalline region and an amorphous cap region. The cylindric region extends in $r$-direction ($\vartheta = \pi/2$) from $r=0$ to $r=r_2$ and in $z$-direction from $z=0$ to $z_\mathrm{a}=\sqrt{r_\text{be}^{2}-r_2^2}$. The amorphous cap region is bounded by the interface with the crystalline region at the bottom $z=z_\mathrm{a}$ and by the hemisphere $r=r_\text{be}$ at the top. Similarly to the previous case, an analytical equation for resistance can be derived by considering thin parallel cylindrical rings and integrating the conductance in the $r$-direction at $\vartheta = \pi/2$. In contrast to the homogeneous case, the conductance of each ring is calculated as a series connection of two conductors, one describing the amorphous part and one the crystalline part. The derivation outlined in the supplementary information section 4 yields

\begin{equation}
  R_\mathrm{dome,inhom} =\frac{\rho_\mathrm{a}}{2\pi \left( r_\mathrm{be}-z_\mathrm{a}\right)}\left[1+\frac{z_\mathrm{a}}{r_\mathrm{be}-z_\mathrm{a}}\left( 1-\frac{\rho_\mathrm{c}}{\rho_\mathrm{a}}\right)\ln{\left( 1+ \frac{\rho_\mathrm{a}}{\rho_\mathrm{c}} \left(  \frac{r_\mathrm{be}}{z_\mathrm{a}}-1\right)\right)}\right]^{-1}
    \label{equ:inhomRdome}
\end{equation}

where $\rho_\text{a}$ and $\rho_\text{c}$ are the amorphous and the crystalline resistivity, respectively. In case the amorphous and crystalline phases are interchanged, the same equation applies, but replacing $\rho_\text{a}$ with $\rho_\text{c}$ and vice versa. The amorphous and crystalline resistivity are not constants, but in the most general form a function of the local temperature and the electric field. Here, we assume that $\rho_\text{c}$ depends only on the temperature according to an Arrhenius-type law, as

\begin{equation}
    \rho_\mathrm{c} =  \rho_\mathrm{c,0}\exp \left( \frac{E_\mathrm{c}}{k_\mathrm{b}T} \right)
    \label{equ:rho_crys}
\end{equation}

In Equation\,(\ref{equ:rho_crys}), $\rho_\mathrm{c,0}$ is the resistivity prefactor of the crystalline phase, $E_\mathrm{c}$ is the activation energy of the carrier transport in the crystalline phase, and $k_\mathrm{b}$ is the Boltzmann constant. For the resistivity $\rho_\mathrm{a}$ of the amorphous phase, a temperature and electric field activated transport (Poole-type) is assumed leading to

\begin{equation}
    \rho_\mathrm{a} =  \rho_\mathrm{a,0}\exp \left( \frac{E_\mathrm{a}-e\beta_0 E_\mathrm{amorph}}{k_\mathrm{b}T} \right)
    \label{equ:rho_amorph}
\end{equation}

In Equation\,(\ref{equ:rho_amorph}), $\rho_\mathrm{a,0}$ is the resistivity prefactor of the amorphous phase, $E_\mathrm{a}$ is the activation energy of the carrier transport in amorphous phase, $E_\mathrm{amorph}$ is the electric field over the amorphous region, and $\beta_0$ is a barrier lowering factor. It should be noted that some studies indicate a Poole-Frenkel-type conduction mechanism rather than a Poole-type \cite{Gallo2015001}. For mathematical simplicity, we used Poole-type in this study. The field dependence of the amorphous phase is only included in the shell resistances $R_\mathrm{a}$ and $R_\mathrm{a,leak}$. In this case, the electric field in radial direction is constant and the integration over the angle $\vartheta$ can be solved analytically. For the heterogeneous dome, the electric field varies along the $r$-direction, which leads to integrals that cannot be solved analytically anymore. As the dome is bypassed by the leakage path, we note that disregarding the field dependence of $R_\mathrm{d}$ only leads to small errors. \\

For the programming model, the drift of the amorphous phase is considered in terms of a power law

\begin{equation}
    \rho_\mathrm{a,0} =  \rho_\mathrm{a,t0}\left( \frac{t}{t_0} \right) ^\nu
    \label{equ:rho_drift}
\end{equation}

where $\rho_\mathrm{a,t0}$ is the resistivity at time $t_0 = 1\,\mathrm{s}$, and $\nu$ is the drift coefficient. \\

The current $I_\mathrm{MSM}$ through the back-to-back diodes in the equivalent circuit diagrams is given by \cite{Sarwat2022001}

\begin{equation}
    I_\mathrm{MSM} =  \frac{2I_\mathrm{D1} I_\mathrm{D2} \sinh \left( \frac{eV_\mathrm{MSM}}{2nk_\mathrm{b}T} \right) }{I_\mathrm{D1} \exp \left( - \frac{eV_\mathrm{MSM}}{2nk_\mathrm{b}T} \right) + I_\mathrm{D2} \exp \left(\frac{eV_\mathrm{MSM}}{2nk_\mathrm{b}T} \right) }
    \label{equ:I_MSM}
\end{equation}
with the current prefactors
\begin{equation}
    I_\mathrm{D1} =  eN\mu \xi_\mathrm{D,I}A_1 \left( u_\mathrm{a} \right) \exp \left(-\frac{e \varphi_\mathrm{el,am}}{k_\mathrm{b}T} \right)
    \label{equ:I_D1}
\end{equation}
and
\begin{equation}
    I_\mathrm{D2} =  eN \mu \xi_\mathrm{D,II}A_2 \left( u_\mathrm{a} \right) \exp \left(-\frac{e \varphi_\mathrm{am,crys}}{k_\mathrm{b}T} \right)
    \label{equ:I_D2}
\end{equation}

In equations (\ref{equ:I_MSM})-(\ref{equ:I_D2}), $V_\mathrm{MSM}$ denotes the added voltage drop over both back-to-back diodes, $n$ is the non-ideality coefficient, $\mu$ is the electron mobility in the PC material, $N$ is the concentration of electrons in the PC material, $\varphi_\mathrm{el,am}$ ($\varphi_\mathrm{am,crys}$) is the barrier height between the heater and the amorphous PC (the amorphous and crystalline PC) material, and $\xi_\mathrm{D,I}$ $\left(\xi_\mathrm{D,II}\right)$ is the surface electric field in the semiconductor at the corresponding interfaces. The areas $A_1$ and $A_2$ describe the area of the interface regions and depend on the size of the amorphous mark $u_\mathrm{a}$. For the leakage path, the two back-to-back diodes approach each other if $u_\mathrm{a}$ approaches $r_\mathrm{be}$. Then, also the depletion zones of the two diodes overlap, and electronic screening lowers the barrier heights. To account for such an effect and avoid current jumps when the diodes disappear at $u_\mathrm{a} = r_\mathrm{be}$, the barriers heights $\varphi_\mathrm{el,am,L}$ and $\varphi_\mathrm{am,cry,L}$ are made dependent on the mark size $u_\mathrm{a}$ according to

\begin{equation}
    \varphi_\mathrm{am,cry,L(el,am,L)} =  \varphi_\mathrm{am,cry (el,am)}\tanh{\left( \frac{u_\mathrm{a}-r_\mathrm{be}}{2.5\,\mathrm{nm}}\right)}
    \label{equ:varphi_L}.
\end{equation}

Applying equations (\ref{equ:Rshell})-(\ref{equ:inhomRdome}) and (\ref{equ:I_MSM})-(\ref{equ:I_D2}), to the circuit elements in \ref{fig:figure2}(d)-(f), inserting equations (\ref{equ:rho_crys})-(\ref{equ:rho_drift}),(\ref{equ:varphi_L}), and applying Kirchhoff's laws leads to a full set of equations allowing us to calculate the resistance of the PCM device at any specific temperature and read voltage for any chosen $u_\mathrm{a} \in \left[0, t_\mathrm{PCM}\right]$, where $t_\mathrm{PCM}$ is the thickness of the PC layer. The assumption of constant temperature is only valid if the read voltage is low enough to avoid any Joule heating. For each of the three circuit diagrams, a specific set of equations is obtained, which has to be chosen according to the size $u_\mathrm{a}$ of the amorphous mark. The resulting set of equations are given in the supplementary information section 5. As the equations are partially implicit, the resulting equation system is solved using a Newton iteration. \\

\begin{figure}
  \includegraphics[width=\linewidth]{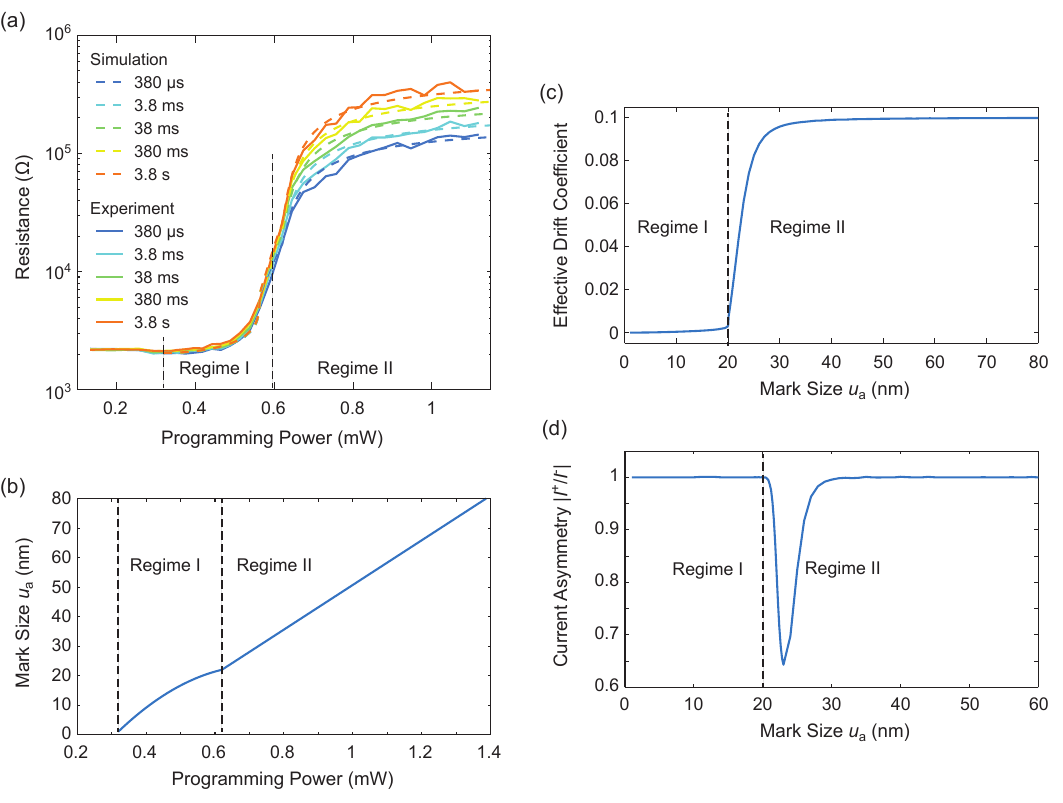}
  \caption{(a) Simulated programming curve (dashed lines) in comparison to experimental data (solid lines) at $T=375 \text{K}$ at five different evaluation times showing the resistance drift behavior. The resistance was determined at $V_\text{read}=0.1\,\text{V}$. (b) Mapping function linking programming power to the amorphous mark size used in the compact model. (c) State-dependent drift coefficient extracted from the data in (a). (d) Current asymmetry of the $I-V$ characteristic determined at $V_\text{read}=\pm 0.4\,\text{V}$}
  \label{fig:figure4}
\end{figure}

To match the analytical model with experimental data, the state variable $u_\mathrm{a}$ of the model must be transformed into a quantity that is experimentally accessible. In a typical PCM programming curve based on amorphization, the resistance is a function of the programming power. The programming power is associated with the size of the amorphous mark, i.e., a larger programming power leads to a larger amorphous mark. To find a mapping between $u_\mathrm{a}$ and the programming power $P_\mathrm{prog}$, we conducted electro-thermal FEM simulations with different programming powers (details are given in the supplementary information section 2). The size of the amorphous mark is defined by the isotherm $T=T_\mathrm{melt}$, where $T_\mathrm{melt}$ is the melting temperature of the PC material (taken as 880\,K, reported values range from 880\,K  to 900\,K \cite{Yamada1991001,Woods2017003}). Three simulated temperature distributions are shown in Fig.\ref{fig:figure1}(b)-(d). The extracted mapping function $u_\mathrm{a} \left( P_\mathrm{prog} \right)$ is shown in Figure\,\ref{fig:figure4}(b). Crucially, the mapping function shows two different regimes. In regime II, the size $u_\mathrm{a}$ of the dome increases linearly with the programming power. In this regime, the heater is fully covered with the amorphous dome. In regime I,  the amorphous mark is within the crystalline volume, and the size depends nonlinearly on the programming power.\\

In this study, Ge$_{2}$Sb$_{2}$Te$_{5}$ (GST) PC material based mushroom-type PCM devices were utilized to validate the model.  In order to ensure stable device operation for experiments, the devices were conditioned with 100,000 RESET programming pulses.  Figure\,\ref{fig:figure4}(a) illustrates programming curves of the PCM device as solid lines in comparison to the simulation model (dashed lines).  The programming curves are obtained by measuring the device resistance at different times from RESET operation. As is evident, the model reproduces not only the shape of the programming curve but also the resistance drift as shown by the different colors. Based on the model, we can divide the programming curve into different regimes. In regime II, the size of the amorphous mark is larger than the radius of the bottom electrode, i.e., $u_\mathrm{a}>r_\mathrm{be}$. Regime I is equivalent to the partial RESET regime. The comparison reveals two interesting results. Already in the partial RESET regime where a fully crystalline path exists, the programmed resistance increases already by a factor of about five. In this regime, the crystalline path is constricted until the heater is fully covered with the amorphous dome. In regime II, first a steep resistance increases appears followed by a more gradual transition with increasing programming power. In the first steep regime, the resistance drift is less pronounced (note that the traces overlap each other) compared to the gradual regime. We note that in the steep regime, the resistance is dominated by the leakage path. The sharp rise in resistance due to the decreasing electric field with increasing $u_\mathrm{a}$ (changes by about three nanometers), which leads to drastic increase of the resistance as the amorphous shell resistance $R_\mathrm{a,leak}$ in the leakage path is field-dependent. \\

From these programming curves, the effective drift coefficient of the PCM device is extracted. A sharp increase of the effective drift coefficient for $20\,\mathrm{nm}<u_\mathrm{a}<30\,\mathrm{nm}$ appears while it stays rather constant for larger sizes of the amorphous mark as shown in Figure\,\ref{fig:figure4}(d). For $20\,\mathrm{nm}<u_\mathrm{a}<30\,\mathrm{nm}$, the resistance of the leakage path (which dominates the overall current) is given by the voltage divider of the amorphous and crystalline shell. Here, there is still a relevant voltage drop over $R_\mathrm{c,leak}$ which decreases with growing $u_\mathrm{a}$. Due to this contribution an effective drift coefficient is extracted that is lower than the nominal one of the amorphous phase. Please note that the drift coefficient of the crystalline phase is set to zero. For non-zero values of the drift coefficient of the amorphous phase, though, the resulting effective drift coefficient would be still a value between the crystalline and amorphous one as determined by the voltage divider in the leakage path. In the proposed model, bulk values of the drift coefficients of the different phases are used and the effective drift coefficient is a result of the voltage divider described by the size and shape of the amorphous mark. This result shows that it is important to account for the size and shape information in an advanced model for PCM mushroom devices.\\

In a further simulation study, the asymmetry of the programmed resistance with respect to the voltage polarity is investigated. Figure\,\ref{fig:figure4}\,(d) shows the read current ratio $\left |I\left(0.4\,\mathrm{V}\right)/I\left (-0.4\,\mathrm{V}\right) \right|$ as a function of the mark size $u_\mathrm{a}$. For low and high values of $u_\mathrm{a}$ the ratio is one, whereas it dips in the range $20\,\mathrm{nm} \leq u_\mathrm{a} \leq 25\,\mathrm{nm}$. This asymmetry results from the inequality of the barriers $\varphi_\mathrm{el,am,L}\left(u_\mathrm{a} \right)$, $\varphi_\mathrm{am,cry,L}\left(u_\mathrm{a} \right)$ and the corresponding diode areas $A_\mathrm{1,L}\left(u_\mathrm{a} \right)$, $A_\mathrm{2,L}\left(u_\mathrm{a} \right)$. For $u_\mathrm{a} \leq r_\mathrm{be}$ the current is dominated by the leakage path. In this regime, the back-to-back diodes are not included in this path, and thus, the current is symmetric. If $u_\mathrm{a} > r_\mathrm{be}$, the amorphous-electrode and amorphous-crystalline interface form electrostatic barriers, which have different barrier heights. As the amorphous-crystalline is higher than the amorphous-electrode one, the former interface dominates when only considering the barrier height. This leads to a potential asymmetry of the electronic transport. This asymmetry builds up according to equation (\ref{equ:varphi_L}) within the first few nanometers. On the other hand, the asymmetry between the areas $A_\mathrm{1,L}\left(u_\mathrm{a} \right)$ and $A_\mathrm{2,L}\left(u_\mathrm{a} \right)$ changes, too.  The area of the amorphous-crystalline interface increases, while the amorphous-electrode interface stays constant when $u_\mathrm{a}$ is increasing. Thus, the amorphous-electrode interface becomes more dominating. These two counteracting effects lead first to a decrease of the ratio (barrier height effect) followed by an increase of the asymmetry (area effect). Finally, the asymmetry vanishes as $R_\mathrm{a,L}$ dominates the total resistance of the leakage path. The dip is consistent with experimental data of PCM data published in literature \cite{Sarwat2022001}.

\section*{An Understanding of Threshold Switching Characteristics}

\begin{figure}
  \includegraphics[width=7.76 cm]{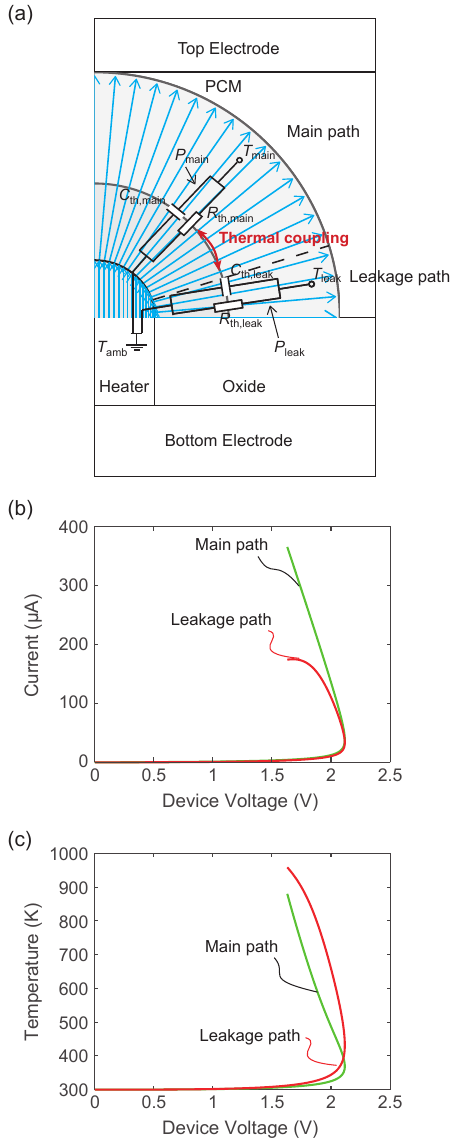}
  \caption{(a) Equivalent circuit diagram of the thermal model of the two current path including a thermal coupling element. (b) Simulated \textit{I-V} curves for the main current path and the leakage current path and  (c) corresponding \textit{T-V} for the two paths.}
  \label{fig:figure5}
\end{figure}

\noindent To extend the analytical model for threshold switching, equations describing the dynamic Joule heating according to equation\,(\ref{equ:Tpcm}) was used. As we consider two current paths, i.e., the main path through the dome and the leakage path, and as the currents via the two paths can differ largely, we introduced two temperatures $T_\mathrm{dome}$ and $T_\mathrm{leak}$, which are described by the equivalent thermal circuit diagram shown in Figure\,\ref{fig:figure5}(a). Each path is described by an effective thermal resistance $R_\mathrm{th,dome}$/$R_\mathrm{th,leak}$ and an equivalent thermal capacitance $C_\mathrm{th,dome}$/$C_\mathrm{th,leak}$. As the two temperatures describe physically adjacent points a thermal coupling is introduced in the model, which is described by a coupling coefficient $\Gamma$. Thus, the dynamic heat equation\,(\ref{equ:Tpcm}) is modified to capture the coupling as

\begin{equation}
   \frac{\mathrm{d}T_\mathrm{dome}}{\mathrm{d}t} =  \frac{R_\mathrm{th,dome}I_\mathrm{dome}V_\mathrm{app}-\left( T_\mathrm{dome}-T_\mathrm{amb} \right) + \Gamma \left(T_\mathrm{leak}-T_\mathrm{dome} \right)}{ R_\mathrm{th,dome}C_\mathrm{th,dome}},
    \label{equ:Tdome}
\end{equation}

and, 

\begin{equation}
   \frac{\mathrm{d}T_\mathrm{leak}}{\mathrm{d}t} =  \frac{R_\mathrm{th,leak}I_\mathrm{leak}V_\mathrm{app}-\left( T_\mathrm{leak}-T_\mathrm{amb} \right) + \Gamma \left(T_\mathrm{dome}-T_\mathrm{leak} \right)}{ R_\mathrm{th,leak}C_\mathrm{th,leak}}.
    \label{equ:Tleak}
\end{equation}

In Equations (\ref{equ:Tdome}) and (\ref{equ:Tleak}), the dissipated power is determined by the corresponding currents $I_\mathrm{dome}$ and $I_\mathrm{leak}$, respectively, multiplied by the device voltage $V_\mathrm{dev}$. These two equations are solved along with the electrical model of the three different shapes. To solve the two ODEs an implicit Euler scheme and an adaptive time stepping method is used. At each time step, the full respective equation system is solved using a Newton iteration. More details are given in the supplementary information section 5.\\

The threshold switching behavior is studied using a voltage sweep of 3\,MV/s with a peak voltage of 3\,V and $u_\mathrm{a}=50\,\mathrm{nm}$ The simulation is terminated when one of the two temperatures reaches the melting temperature $T_\mathrm{melt}=880\,\mathrm{K}$ of the PC material. Figure\,\ref{fig:figure5}(b) and (c) show the simulated $I-V$ and $T-V$ characteristics, respectively. The currents are plotted versus the device voltage $V_\mathrm{dev} = V_\mathrm{app}-I_\mathrm{PCM}R_\mathrm{s}$. The temperature increases first in the leakage path that leads to the thermally-induced threshold switching. Consequently, due to the thermal coupling, the temperature in the dome increases after some delay as well. This shows the importance of the leakage path for the threshold switching. Within our proposed model, the threshold switching is determined by the leakage path, and thus, the determining electric field is defined by the voltage drop over $R_\mathrm{a,leak}$ divided by the length of the amorphous region $u_\mathrm{a}-r_\mathrm{be}$. Notably, this definition of the electric field in order to describe the critical field for threshold switching has been proposed in literature \cite{Gallo2016001}, albeit, without a physical attribution. The proposed model shows that this critical field results directly from the leakage path that has distinct properties. Furthermore, the continuum model simulations of PCM mushroom devices have shown that the hottest spot is expected close to the injection point of the leakage path\cite{Gallo2016001}, which is consistent with our model.\\

\begin{figure}
  \includegraphics[width=\linewidth]{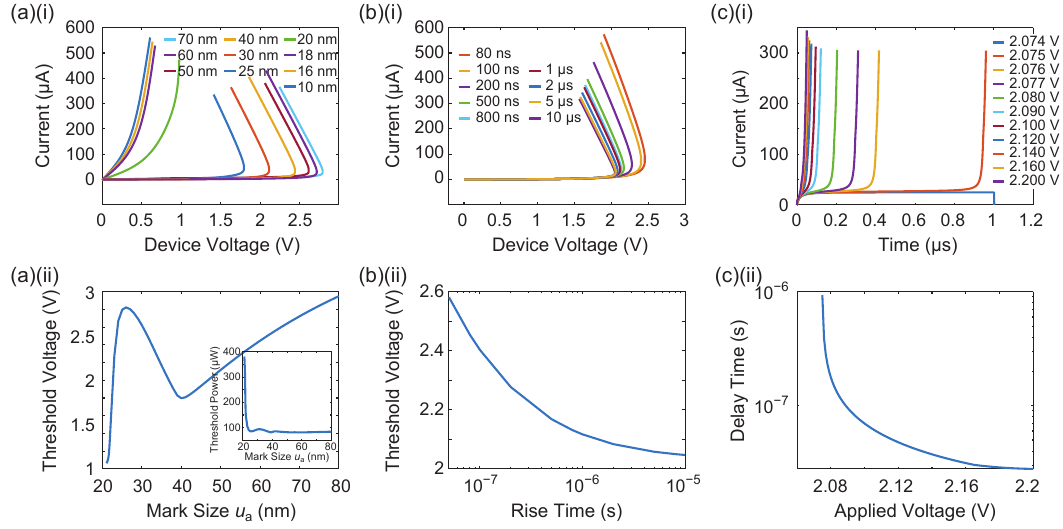}
  \caption{Influence of (a) the programmed state, (b) the sweep rate, and (c) the square pulse voltage on the threshold
switching characteristics. (a) (i) and (ii) show the \textit{I-V} characteristics and the threshold voltages $V_\text{th}$ as a function of the read resistance and $u_\text{a}$, respectively. The dependence of the \textit{I-V} characteristics and the threshold voltages on the sweep rate are shown in (b) (i) and (ii). (c) (i) and (ii) show the \textit{I-t} characteristics for different pulse voltages and their corresponding \textit{t-V} characteristics.}

  \label{fig:figure6}
\end{figure}

To further build confidence on the proposed model, several parameter studies are conducted, and their influence on the threshold voltage is evaluated. The threshold voltage is extracted as the maximum voltage $V_\mathrm{dev}$ of the snapback \textit{I-V} characteristic. First, the size of the amorphous mark was changed to investigate its influence on the threshold voltage. To this end $u_\mathrm{a}$ was varied from 0\,nm-80\,nm, while the voltage sweep rate was kept at 3\,MV/s and the reference temperature was set to $T_\mathrm{ref}=300\,\mathrm{K}$, i.e, slightly increased ambient temperature. Figure\,\ref{fig:figure6}(a)(i) shows the simulated $I-V_\mathrm{dev}$ characteristics for selected $u_\mathrm{a}$. Two different regimes evolve. If $u_\mathrm{a}>r_\mathrm{be}$ clear threshold switching characteristics showing a negative differential resistance (NDR) region are observed. In contrast, no NDR appears for $u_\mathrm{a} < r_\mathrm{be}$. In the former regime, the threshold voltage increases with the size of the amorphous mark (RESET resistance extracted at $V_\mathrm{read}=0.2\,\mathrm{V}$) as illustrated in Figure\,\ref{fig:figure6}(a)(ii). For $u_\mathrm{a} = 20\,\mathrm{nm}-40\,\mathrm{nm}$, the threshold voltage shows a peak with a maximum at $24\,\text{nm}$. In this regime, there is strong influence of the field-dependent conductivity in the leakage path, which leads to complex change of the voltage divider between the amorphous and crystalline part of the leakage path. For larger amorphous marks, the threshold voltage increases approximately linearly, which is qualitatively consistent with the experimental observations of various groups \cite{Gallo2016001,Pigot2018002}. The inset of Figure\,\ref{fig:figure6}(a)(ii) shows the threshold power as a function of the size of the amorphous mark. The threshold power is almost constant for amorphous marks larger than about $22\,\text{nm}$ preceded by a sharp decrease in threshold power. In the latter regime, there is a significant leakage current, as the diodes are not fully established yet.\\

In the next simulation study, the rise time of the voltage sweep is varied from 80\,ns-10\,µs. Here, we used an amorphous size of $u_\mathrm{a}=50\,\mathrm{nm}$ to stay in a clear threshold switching regime. As shown in Figures\,\ref{fig:figure6}(b)(i) and (ii), while the NDR effect is observed in all simulated rise times, the threshold voltage increases with decreasing rise time. This trend is also consistent with experimental observations \cite{Gallo2016001}. This behavior is further verified in another parameter study. Here, the delay in threshold switching is investigated using rectangular voltage pulses with varying amplitude. The rise time of the voltage pulse is 1\,µs. The delay time of the threshold switching is defined as the time at which a threshold current of 60\,µA is reached. As shown in Figure\,\ref{fig:figure6}(c)(i) and (ii), the delay time is very sensitive to a change of the voltage amplitude. During the course of the voltage pulse, the current shows first a plateau-like region and then a sharp switching event. For increasing voltages, the plateau-like region gets smaller and finally vanishes. This observation has also been made in the literature \cite{Gallo2016001,Saxena2019001,Saxena2021001}. Together, these affects can be ascribed to the electronic transition in the multiple-trapping picture of transport in PC materials: the delay time in the rise of material conductivity is a result of the thermal time constant of the system defined by the values of the thermal capacitances and the thermal resistances. \\

With these additional examples, we can faithfully establish that the proposed compact model is capable of capturing the key characteristics of the devices for both programming and read-out behaviors. Note that while we have demonstrated this for a standard mushroom-type device, it is also straightforward to extend the compact model to projected-type devices\cite{syed2023memory}. This would include accounting for the resistance of the projecting layer in the lateral direction bypassing the amorphous dome and the resistance of the projecting layer in the vertical direction in series to the dome resistance. Also note that although our descriptions have primarily focused on mushroom-type devices, the equation sets are applicable to other PCM device geometries as well. For example, similar effects can be expected in commercial wall-structure type heater-based PCM devices. Because the heater spans one entire dimension in such as device, the field enhancement effects might be even more significant. Since the phase configurations are also dome-shaped, adapting the resistor network description should allow modelling of the programming and read-out characteristics for these geometries as well\cite{Y2010Sandre,boniardi2014optimization}. Note that while continuum models based on finite element methods can capture the full three-dimensional shape of the amorphous region and are typically used to model programming characteristics, they are computationally intensive and not suitable for circuit-level simulations. This is where the proposed modeling framework makes a significant contribution.

\section*{Conclusion}

\noindent In summary, we have developed a compact model that captures both the programming and read-out characteristics of mushroom-type PCM devices. A key feature of the model is its ability to account for the shape and size of the amorphous mark, allowing key experimental observations to emerge naturally from the formulation. The model also leads to an equivalent electrical circuit representation that includes a leakage path. This path represents current injection at the rim of the heater structure into the PC material, and it enables accurate reproduction of both the programming curve and threshold switching behavior. Furthermore, the model effectively captures state-dependent variations in the drift coefficient as well as voltage polarity–dependent current asymmetry. The model is compatible with standard circuit simulation tools, and a Verilog-A implementation is provided to simulate the devices at the circuit level.

\section*{Methods}

\small

\section*{Experimental Section}

\noindent A custom-made electrical characterization platform with integrated heating stage was used for electrical measurements. The sample was mounted on an invar block with two embedded tungsten heaters. The temperature was measured using a thermocouple inserted into the invar block and controlled via a Eurotherm 2416 temperature controller. The drift experiment is performed with an Agilent 81150A Pulse Function Arbitrary Generator that allows to combine the signal of two internal independent pulse generators. One generator sends the RESET pulse and the second a burst of triangular read pulses (\unit[200]{mV}). The transient voltage signal and device current are measured with an oscilloscope. The current is amplified with an operational amplifier circuit. Each pulse of the burst gives the device’s IV characteristic. By fitting those, the time-resolved device resistance is obtained.

\section*{Modelling Section}
\noindent The analytical model was developed using MATLAB R2019 and the Verilog-A file was verified using CADENCE SPECTRE.

% \section*{Supporting Information} 
% Supporting Information is available from the Wiley Online Library. 

\section*{Acknowledgment} %delete if not applicable))
\noindent This work was supported by the IBM Research AI Hardware Center. The work is partially funded by the BMBF project Neurotec 2 under the Grant 16ME0398K.

\section*{Contributions}

\noindent S.M. developed the analytical models and simulation frameworks. B.K. contributed electrical device measurements. R.W.A. contributed to developing the Verilog-A model. A.S. provided essential input and management support. G.S.S. defined the research question and provided supervision. S.M. and G.S.S. wrote the manuscript with input from all authors.

\section*{References}
\bibliographystyle{naturemag}
\bibliography{ProgModel}

\begin{thebibliography}{10}
\expandafter\ifx\csname url\endcsname\relax
  \def\url#1{\texttt{#1}}\fi
\expandafter\ifx\csname urlprefix\endcsname\relax\def\urlprefix{URL }\fi
\providecommand{\bibinfo}[2]{#2}
\providecommand{\eprint}[2][]{\url{#2}}

\bibitem{syed2025phase}
\bibinfo{author}{Syed, G.~S.}, \bibinfo{author}{Le~Gallo, M.} \&
  \bibinfo{author}{Sebastian, A.}
\newblock \bibinfo{title}{Phase-change memory for in-memory computing}.
\newblock \emph{\bibinfo{journal}{Chemical Reviews}}
  \textbf{\bibinfo{volume}{125}} (\bibinfo{year}{2025}).

\bibitem{Burr2016003}
\bibinfo{author}{Burr, G.~W.} \emph{et~al.}
\newblock \bibinfo{title}{Recent progress in phase-change memory technology}.
\newblock \emph{\bibinfo{journal}{IEEE J. Emerging Sel. Top. Circuits Syst.}}
  \textbf{\bibinfo{volume}{6}}, \bibinfo{pages}{146--162}
  (\bibinfo{year}{2016}).

\bibitem{Ehrmann2022001}
\bibinfo{author}{Ehrmann, A.}, \bibinfo{author}{Blachowicz, T.},
  \bibinfo{author}{Ehrmann, G.} \& \bibinfo{author}{Grethe, T.}
\newblock \bibinfo{title}{Recent developments in phase-change memory}.
\newblock \emph{\bibinfo{journal}{Applied Research}}
  \textbf{\bibinfo{volume}{1}} (\bibinfo{year}{2022}).

\bibitem{Gallo2020001}
\bibinfo{author}{Gallo, M.~L.} \& \bibinfo{author}{Sebastian, A.}
\newblock \bibinfo{title}{An overview of phase-change memory device physics}.
\newblock \emph{\bibinfo{journal}{J. Phys. D Appl. Phys.}}
  \textbf{\bibinfo{volume}{53}}, \bibinfo{pages}{213002}
  (\bibinfo{year}{2020}).

\bibitem{Nardone2012001}
\bibinfo{author}{Nardone, M.}, \bibinfo{author}{Simon, M.},
  \bibinfo{author}{Karpov, I.~V.} \& \bibinfo{author}{Karpov, V.~G.}
\newblock \bibinfo{title}{Electrical conduction in chalcogenide glasses of
  phase change memory}.
\newblock \emph{\bibinfo{journal}{J. Appl. Phys.}}
  \textbf{\bibinfo{volume}{112}}, \bibinfo{pages}{071101}
  (\bibinfo{year}{2012}).

\bibitem{Kaes2015001}
\bibinfo{author}{Kaes, M.}, \bibinfo{author}{Gallo, M.~L.},
  \bibinfo{author}{Sebastian, A.}, \bibinfo{author}{Salinga, M.} \&
  \bibinfo{author}{Krebs, D.}
\newblock \bibinfo{title}{High-field electrical transport in amorphous
  phase-change materials}.
\newblock \emph{\bibinfo{journal}{J. Appl. Phys.}}
  \textbf{\bibinfo{volume}{118}}, \bibinfo{pages}{135707/1--11}
  (\bibinfo{year}{2015}).

\bibitem{Sarwat2022001}
\bibinfo{author}{Sarwat, S.~G.} \emph{et~al.}
\newblock \bibinfo{title}{Mechanism and impact of bipolar current voltage
  asymmetry in computational phase-change memory (early view)}.
\newblock \emph{\bibinfo{journal}{Adv. Mater.}} \bibinfo{pages}{2201238}
  (\bibinfo{year}{2022}).

\bibitem{Gallo2015001}
\bibinfo{author}{Gallo, M.~L.}, \bibinfo{author}{Kaes, M.},
  \bibinfo{author}{Sebastian, A.} \& \bibinfo{author}{Krebs, D.}
\newblock \bibinfo{title}{Subthreshold electrical transport in amorphous
  phase-change materials}.
\newblock \emph{\bibinfo{journal}{New Journal of Physics}}
  \textbf{\bibinfo{volume}{17}}, \bibinfo{pages}{093035}
  (\bibinfo{year}{2015}).

\bibitem{Zhang2019008}
\bibinfo{author}{Zhang, W.}, \bibinfo{author}{Mazzarello, R.},
  \bibinfo{author}{Wuttig, M.} \& \bibinfo{author}{Ma, E.}
\newblock \bibinfo{title}{Designing crystallization in phase-change materials
  for universal memory and neuro-inspired computing}.
\newblock \emph{\bibinfo{journal}{Nat Rev Mater}} \textbf{\bibinfo{volume}{4}},
  \bibinfo{pages}{150--168} (\bibinfo{year}{2019}).

\bibitem{Sebastian2014001}
\bibinfo{author}{Sebastian, A.}, \bibinfo{author}{Gallo, M.~L.} \&
  \bibinfo{author}{Krebs, D.}
\newblock \bibinfo{title}{Crystal growth within a phase change memory cell}.
\newblock \emph{\bibinfo{journal}{Nature Communications}}
  \textbf{\bibinfo{volume}{5}}, \bibinfo{pages}{4314/1--}
  (\bibinfo{year}{2014}).

\bibitem{Menzel2015001}
\bibinfo{author}{Menzel, S.}, \bibinfo{author}{Böttger, U.},
  \bibinfo{author}{Wimmer, M.} \& \bibinfo{author}{Salinga, M.}
\newblock \bibinfo{title}{Physics of the switching kinetics in resistive
  memories}.
\newblock \emph{\bibinfo{journal}{Adv. Funct. Mater.}}
  \textbf{\bibinfo{volume}{25}}, \bibinfo{pages}{6306--6325}
  (\bibinfo{year}{2015}).

\bibitem{Gallo2016001}
\bibinfo{author}{Gallo, M.~L.}, \bibinfo{author}{Athmanathan, A.},
  \bibinfo{author}{Krebs, D.} \& \bibinfo{author}{Sebastian, A.}
\newblock \bibinfo{title}{Evidence for thermally assisted threshold switching
  behaviour in nanoscale phase-change memory cells}.
\newblock \emph{\bibinfo{journal}{J. Appl. Phys.}}
  \textbf{\bibinfo{volume}{119}}, \bibinfo{pages}{025704}
  (\bibinfo{year}{2016}).

\bibitem{Wimmer2014001}
\bibinfo{author}{Wimmer, M.} \& \bibinfo{author}{Salinga, M.}
\newblock \bibinfo{title}{The gradual nature of threshold switching}.
\newblock \emph{\bibinfo{journal}{New Journal of Physics}}
  \textbf{\bibinfo{volume}{16}}, \bibinfo{pages}{113044}
  (\bibinfo{year}{2014}).

\bibitem{Kim2011030}
\bibinfo{author}{Kim, S.} \emph{et~al.}
\newblock \bibinfo{title}{Resistance and threshold switching voltage drift
  behavior in phase-change memory and their temperature dependence at
  microsecond time scales studied using a micro-thermal stage}.
\newblock \emph{\bibinfo{journal}{IEEE Trans. Electron Devices}}
  \textbf{\bibinfo{volume}{58}}, \bibinfo{pages}{584--592}
  (\bibinfo{year}{2011}).

\bibitem{Kersting2020001}
\bibinfo{author}{Kersting, B.} \emph{et~al.}
\newblock \bibinfo{title}{State dependence and temporal evolution of resistance
  in projected phase change memory}.
\newblock \emph{\bibinfo{journal}{Sci. Rep.}} \textbf{\bibinfo{volume}{10}},
  \bibinfo{pages}{8248} (\bibinfo{year}{2020}).

\bibitem{Rizzi2014001002}
\bibinfo{author}{Rizzi, M.} \emph{et~al.}
\newblock \bibinfo{title}{Cell-to-cell and cycle-to-cycle retention statistics
  in phase-change memory arrays}.
\newblock \emph{\bibinfo{journal}{IEEE Trans. Electron Devices}}
  \bibinfo{pages}{2205--11} (\bibinfo{year}{2015}).

\bibitem{Tabatabaei2017001}
\bibinfo{author}{Tabatabaei, F.}, \bibinfo{author}{Boussinot, G.},
  \bibinfo{author}{Spatschek, R.}, \bibinfo{author}{Brener, E.~A.} \&
  \bibinfo{author}{Apel, M.}
\newblock \bibinfo{title}{Phase field modeling of rapid crystallization in the
  phase-change material aist}.
\newblock \emph{\bibinfo{journal}{J. Appl. Phys.}}
  \textbf{\bibinfo{volume}{122}}, \bibinfo{pages}{045108}
  (\bibinfo{year}{2017}).

\bibitem{Ciocchini2015001}
\bibinfo{author}{Ciocchini, N.} \emph{et~al.}
\newblock \bibinfo{title}{Impact of thermoelectric effects on phase change
  memory characteristics}.
\newblock \emph{\bibinfo{journal}{IEEE Trans. Electron Devices}}
  \textbf{\bibinfo{volume}{62}}, \bibinfo{pages}{3264 -- 3271}
  (\bibinfo{year}{2015}).

\bibitem{Woods2017003}
\bibinfo{author}{Woods, Z.} \& \bibinfo{author}{Gokirmak, A.}
\newblock \bibinfo{title}{Modeling of phase-change memory: Nucleation, growth,
  and amorphization dynamics during set and reset: Part i—effective media
  approximation}.
\newblock \emph{\bibinfo{journal}{IEEE Trans. Electron Devices}}
  \textbf{\bibinfo{volume}{64}}, \bibinfo{pages}{4466--4471}
  (\bibinfo{year}{2017}).

\bibitem{Woods2017004}
\bibinfo{author}{Woods, Z.}, \bibinfo{author}{Scoggin, J.},
  \bibinfo{author}{Cywar, A.}, \bibinfo{author}{Adnane, L.} \&
  \bibinfo{author}{Gokirmak, A.}
\newblock \bibinfo{title}{Modeling of phase-change memory: Nucleation, growth,
  and amorphization dynamics during set and reset: Part ii—discrete grains}.
\newblock \emph{\bibinfo{journal}{IEEE Trans. Electron Devices}}
  \textbf{\bibinfo{volume}{64}}, \bibinfo{pages}{4472--4478}
  (\bibinfo{year}{2017}).

\bibitem{Pigot2018001}
\bibinfo{author}{Pigot, C.} \emph{et~al.}
\newblock \bibinfo{title}{Comprehensive phase-change memory compact model for
  circuit simulation}.
\newblock \emph{\bibinfo{journal}{IEEE Trans. Electron Devices}}
  \textbf{\bibinfo{volume}{65}}, \bibinfo{pages}{4282--4289}
  (\bibinfo{year}{2018}).

\bibitem{Pigot2018002}
\bibinfo{author}{Pigot, C.} \emph{et~al.}
\newblock \bibinfo{title}{Phase-change memory: A continuous multilevel compact
  model of subthreshold conduction and threshold switching}.
\newblock \emph{\bibinfo{journal}{Jpn. J. Appl. Phys.}}
  \textbf{\bibinfo{volume}{57}}, \bibinfo{pages}{04FE13}
  (\bibinfo{year}{2018}).

\bibitem{Ding2022001}
\bibinfo{author}{Ding, F.} \emph{et~al.}
\newblock \bibinfo{title}{A review of compact modeling for phase change
  memory}.
\newblock \emph{\bibinfo{journal}{J. Semicond.}} \textbf{\bibinfo{volume}{43}},
  \bibinfo{pages}{023101} (\bibinfo{year}{2022}).

\bibitem{Kersting2021001}
\bibinfo{author}{Kersting, B.} \emph{et~al.}
\newblock \bibinfo{title}{Measurement of onset of structural relaxation in
  melt-quenched phase change materials}.
\newblock \emph{\bibinfo{journal}{Adv. Funct. Mater.}}
  \textbf{\bibinfo{volume}{n/a}}, \bibinfo{pages}{2104422}
  (\bibinfo{year}{2021}).

\bibitem{Yamada1991001}
\bibinfo{author}{Yamada, N.}, \bibinfo{author}{Ohno, E.},
  \bibinfo{author}{Nishiuchi, K.}, \bibinfo{author}{Akahira, N.} \&
  \bibinfo{author}{Takao, M.}
\newblock \bibinfo{title}{Rapid‐phase transitions of gete‐sb$_2$te$_3$
  pseudobinary amorphous thin films for an optical disk memory}.
\newblock \emph{\bibinfo{journal}{J. Appl. Phys.}}
  \textbf{\bibinfo{volume}{69}}, \bibinfo{pages}{2849–2856}
  (\bibinfo{year}{1991}).

\bibitem{Saxena2019001}
\bibinfo{author}{Saxena, N.} \& \bibinfo{author}{Manivannan, A.}
\newblock \bibinfo{title}{Threshold switching dynamics of pseudo-binary
  {GeTe}{\textendash}sb2te3 phase change memory devices}.
\newblock \emph{\bibinfo{journal}{J. Phys. D Appl. Phys.}}
  \textbf{\bibinfo{volume}{52}}, \bibinfo{pages}{375301}
  (\bibinfo{year}{2019}).

\bibitem{Saxena2021001}
\bibinfo{author}{Saxena, N.}, \bibinfo{author}{Raghunathan, R.} \&
  \bibinfo{author}{Manivannan, A.}
\newblock \bibinfo{title}{A scheme for enabling the ultimate speed of threshold
  switching in phase change memory devices}.
\newblock \emph{\bibinfo{journal}{Sci. Rep.}} \textbf{\bibinfo{volume}{11}},
  \bibinfo{pages}{6111} (\bibinfo{year}{2021}).

\bibitem{syed2023memory}
\bibinfo{author}{Syed, G.} \emph{et~al.}
\newblock \bibinfo{title}{In-memory compute chips with carbon-based projected
  phase-change memory devices}.
\newblock In \emph{\bibinfo{booktitle}{2023 International Electron Devices
  Meeting (IEDM)}}, \bibinfo{pages}{1--4} (\bibinfo{organization}{IEEE},
  \bibinfo{year}{2023}).

\bibitem{Y2010Sandre}
\bibinfo{author}{De~Sandre, G.} \emph{et~al.}
\newblock \bibinfo{title}{A 90nm 4mb embedded phase-change memory with 1.2v
  12ns read access time and 1mb/s write throughput}.
\newblock In \emph{\bibinfo{booktitle}{2010 IEEE International Solid-State
  Circuits Conference - (ISSCC)}}, \bibinfo{pages}{268--269}
  (\bibinfo{year}{2010}).

\bibitem{boniardi2014optimization}
\bibinfo{author}{Boniardi, M.} \emph{et~al.}
\newblock \bibinfo{title}{Optimization metrics for phase change memory (pcm)
  cell architectures}.
\newblock In \emph{\bibinfo{booktitle}{2014 IEEE International Electron Devices
  Meeting}}, \bibinfo{pages}{29--1} (\bibinfo{organization}{IEEE},
  \bibinfo{year}{2014}).

\end{thebibliography}

\newpage

%\section*{Figures}

\end{document}